\documentclass[
  twoside,
  twocolumn,
  prb,
  preprintnumbers,
  floatfix,
  showpacs,
  superscriptaddress,
]{revtex4-1}

\usepackage{amsmath}
\usepackage{amssymb}
\usepackage{color}
\usepackage{graphicx}
\usepackage{tabularx}
\usepackage{subfigure}
\usepackage{dcolumn}
\usepackage{textcomp}
\usepackage{units}
\usepackage{gensymb}
\usepackage{times}
\usepackage{enumerate}

\newcommand{\PdN}[0]{\text{PdN}$_2$}
\newcommand{\OsN}[0]{\text{OsN}$_2$}
\newcommand{\IrN}[0]{\text{IrN}$_2$}
\newcommand{\PtN}[0]{\text{PtN}$_2$}

\newcommand{\Ag}[0]{A$_{\text{g}}$}
\newcommand{\Eg}[0]{E$_{\text{g}}$}

\newcommand{\Tg}[0]{T$_{\text{g}}$}
\newcommand{\braket}[3]{\bigl{<}#1\big|#2\big|#3\bigr{>}}

\newcommand{\ibraket}[2]{\bigl{<}#1\big|#2\bigr{>}}

\newcommand{\sect}[1]{Sec.~\ref{#1}}
\newcommand{\fig}[1]{Fig.~\ref{#1}}
\newcommand{\eq}[1]{Eq.~(\ref{#1})}

\renewcommand{\epsilon}[0]{\varepsilon}

\definecolor{orange}{rgb}{0.133, 0.545, 0.133 }

\newlength{\myhgt}

\begin{document}

\preprint{published in Phys. Rev. B {\bf 82}, 104116 (2010) Copyright (2010) by the American Physical Society}

\title{
Pressure-induced phase transition in the electronic structure of palladium nitride
}
\author{Daniel\,\AA{}berg}
\email{aberg2@llnl.gov}
\affiliation{
  Lawrence Livermore National Laboratory,
  Physical and Life Sciences Directorate,
  Livermore, California 94551, USA
}
\author{Paul Erhart}
\affiliation{
  Lawrence Livermore National Laboratory,
  Physical and Life Sciences Directorate,
  Livermore, California 94551, USA
}
\author{Jonathan Crowhurst}
\affiliation{
  Lawrence Livermore National Laboratory,
  Physical and Life Sciences Directorate,
  Livermore, California 94551, USA
}
\author{Joseph M. Zaug}
\affiliation{
  Lawrence Livermore National Laboratory,
  Physical and Life Sciences Directorate,
  Livermore, California 94551, USA
}
\author{Alexander F. Goncharov}
\affiliation{
  Geophysical Laboratory,
  Carnegie Institution of Washington,
  5251 Broad Branch Road NW,
  Washington, DC 20015, USA
}
\author{Babak Sadigh}
\affiliation{
  Lawrence Livermore National Laboratory,
  Physical and Life Sciences Directorate,
  Livermore, California 94551, USA
}

\begin{abstract}
We present a combined theoretical and experimental study of the electronic structure and equation of state (EOS) of crystalline PdN$_2$. The compound forms above 58\,GPa in the pyrite structure and is metastable down to 11\,GPa. We show that the EOS cannot be accurately described within either the local density or generalized gradient approximations. The Heyd-Scuseria-Ernzerhof exchange-correlation functional (HSE06), however, provides very good agreement with experimental data. We explain the strong pressure dependence of the Raman intensities in terms of a similar dependence of the calculated band gap, which closes just below 11\,GPa. At this pressure, the HSE06 functional predicts a first-order isostructural transition accompanied by a pronounced elastic instability of the longitudinal-acoustic branches that provides the mechanism for the experimentally observed decomposition. Using an extensive Wannier function analysis, we show that the structural transformation is driven by a phase transition of the electronic structure, which is manifested by a discontinuous change in the hybridization between Pd-$d$ and N-$p$ electrons as well as a conversion from single to triple bonded nitrogen dimers. We argue for the possible existence of a critical point for the isostructural transition, at which massive fluctuations in both the electronic as well as the structural degrees of freedom are expected.
\end{abstract}


\maketitle

\section{Introduction}

Solid compounds of some of the platinum metals (PM) and nitrogen have recently been synthesized under high pressure and temperature.\cite{GreSan04, YouSan06, CroGon08, CroGon06a} \footnote{We note that compounds have also been formed using different techniques, e.g., with Pt (Ref.~\onlinecite{soto04}), Rh (Ref.~\onlinecite{MorDi07}), and also with Au (Ref.~\onlinecite{SilHun02}).} These compounds are of fundamental interest and possibly technological importance due to their unusual or extreme properties. \IrN, for example, has a bulk modulus of 428\,GPa and is conceivably superhard. \cite{YouSan06} Experimental and theoretical results \cite{CroGon06a, YouMon06, YuZha07, Aberg08} indicate that the crystal structures of PM-nitrides depend on the group number of the parent metal. Specifically, it has been proposed that the nitrides of Os and Ru crystallize in the marcasite structure, \cite{YuZha07,Aberg08} Rh \cite{YuZha07, Aberg08} and Ir (Refs. \onlinecite{CroGon08, Aberg08, YuZha07}) in the marcasite and baddeleyite structures, respectively, and Pt (Refs. \onlinecite{CroGon06a}, \onlinecite{YouMon06}, and \onlinecite{Yuzha06, vApLum06, Aberg08}) and Pd, \cite{CroGon08,Aberg08,CheTse10} in the pyrite structure. First principles calculations have suggested alternative low-enthalpy structures that have not yet been experimentally observed. \cite{Aberg08, TriZun08, ZhaTriZun09, LiWan09}

One common denominator of all platinum metal nitrides studied to date is that compound formation leads to a large increase in the separation between the transition-metal ions compared to the pure metal phases, which can lead to localization of the transition-metal $d$ electrons at low pressures. In the case of \PtN, \IrN, and \OsN\, the $5d$ electrons are extended to the degree that significant hybridization with N $2p$ electrons occurs at all pressures. As a result, the behavior of these compounds under pressure is uncomplicated; their bonding characteristics as well as optical properties remain qualitatively unchanged. They are recoverable to ambient conditions after synthesis at high pressure and temperature. In stark contrast, the nominal \PdN\ compound does not appear to be recoverable \cite{CroGon08} and, as will be shown in this paper, exhibits strongly pressure-dependent Raman activity. First-principles calculations suggest that both \PtN\ and \PdN\ are metastable at low pressures, \cite{Aberg08} yet only \PtN\ has been recovered to ambient conditions. Enthalpy calculations alone thus cannot explain the decomposition of \PdN. Bearing in mind that the atomic $4d$ wave functions of Pd have a shorter range than their $5d$ counterparts in, e.g., \PtN, the observed anomalies could result from electron localization. It is well known that electron localization is the driving force for highly correlated electronic behavior, leading to, e.g., high-$T$ superconductivity as well as metal-insulator transitions. This kind of behavior is most easily seen in $3d$-transition-metal compounds.

In this paper, we present experimental Raman intensities which are strongly pressure dependent. We use electronic-structure calculations to show that the observed anomalies of the \PdN\ compound are related to changes in hybridization between Pd-$4d$ and N-$2p$ electrons.  We find that the standard exchange-correlation (XC) functionals within density-functional theory, which in the past have so accurately described the $5d$ platinum metal compounds, fail to reproduce the experimental observations. By resorting to a hybrid XC functional that provides a much improved description of localization, we show that it is possible to explain the origin of the apparent decomposition by an isostructural phase transformation, corresponding to the conversion from triply to singly bonded nitrogen dimers that is accompanied by a pronounced long wavelength phonon instability. \cite{YouMon06,CheTse10} We furthermore provide arguments for the existence of a critical point associated with this transition, where strong electronic as well as ionic fluctuations should occur. 

This paper is organized as follows. In \sect{sec:meth} we discuss the experimental and computational methods used in this paper. The resulting Raman spectra and calculated electronic structure as a function of atomic volume, pressure, and internal coordinates are addressed in \sect{sec:results}. The implications of our findings are discussed in \sect{sec:discussion}.

\section{methodology}
\label{sec:meth}
\subsection{Experimental details}

Palladium nitride was synthesized under high pressure and temperature using a laser-heated diamond-anvil cell (DAC). 
Raman measurements were carried out using a Jobin Yvon HR460 grating spectrometer and a Roper Scientific liquid N$_2$ cooled charge coupled device (CCD) detector. Spectra were acquired using a grating of 1800\,lines/mm, providing a nominal resolution of 1\,cm$^{-1}$. 
No attempt was made to correct for instrumental nonlinearity as this was assumed to be unimportant over the frequency range of interest. A continuous wave linearly polarized Ar ion laser operating at 488\,nm  was used to excite and measure the Raman spectra. \footnote{The two highest pressure points of experimental Run 1 were obtained with an Ar ion laser operating at 458\,nm, an HR460 spectrometer and a Princeton Instruments CCD detector.} The laser was focused onto the sample using a Mitutoyo 20$\times$ apochromatic near-infrared corrected objective providing a focal spot on the order of 1\, $\mu$m. A continuous wave Nd:YAG laser operating at 1064\,nm that was aligned collinearly with the probe laser was used to heat the reactants. Diamond-anvil cells equipped with diamonds of culet dimension 300 $\mu$m and rhenium gaskets were used to generate the necessary pressure. Pressures were determined according to the ruby fluorescence scale of Mao. \cite{MaoBelSha78} Ruby chips within 20 $\mu$m of the probed area of the sample were used for this purpose and yielded pressures that we estimate to be accurate to within 2\,GPa. 

Raman spectra were obtained in two experimental runs. For Run 1 nitrogen (Air Products, ultrahigh purity grade) was loaded at room temperature under high pressure into the DAC together with the Pd metal while for Run 2 it was loaded as a cryogenic liquid (99.998\%, Praxair). The Pd metal was the same in both cases (99.95\%, 200 mesh, Alfa Aesar, 12068). Pd metal was thermally insulated from one diamond by a thin layer of ruby.  Before introducing liquid nitrogen into the DAC cavity the entire DAC was placed in a dry-nitrogen purged glove box and heated at 383\,K for 12\,h after which the sample chamber was sealed. The DAC was then removed from the glove box, placed in an open-mouth dewar and rapidly immersed in liquid nitrogen. The dewar mouth was covered to generate a positive N$_2$ gas head pressure to discourage atmospheric contamination (e.g., by  H$_2$O or O$_2$). A few minutes later the sample chamber was opened to introduce liquid nitrogen and then closed and pressurized. \footnote{Using less pure starting materials and less controlled procedures resulted in the appearance of a strong Raman spectrum upon heating at low pressure that was very likely due to PdO which is known to be resonantly enhanced. We also observed other features not accounted for by our theoretical calculations and possibly due to impurities, see discussion and Fig. 1 of Ref.\, \onlinecite{CroGon08}.}

\subsection{Computational details}
All first principles calculations, excluding the Raman intensities, used the projector augmented wave (PAW) method \cite{Blo94} as implemented in the Vienna {\em ab initio} simulation package (\textsc{vasp}) \cite{KreFur96a} and the Heyd-Scuseria-Ernzerhof XC functional (HSE06). \cite{HeyScuErn03} Comparisons are also made with the local-density approximation (LDA) and generalized gradient approximation (GGA) as parametrized by Perdew, Burke, and Ernzerhof. \cite{PerBurErn96} The Brillouin zone integrations were performed using $8\times8\times8$ and $6\times6\times6$ Monkhorst-Pack $k$-point grids \cite{MonPac76} for structural relaxations and calculation of Raman frequencies, respectively. The plane-wave cutoff was set to 520\, eV, and Gaussian smearing with a width of 0.1\, eV was used to determine the occupation numbers. Atomic positions were relaxed with ionic forces converged to 20\, meV/\AA.

\subsubsection{Raman intensities}
In a Raman-scattering experiment, the sample is exposed to a monochromatic light source. Photons with angular frequency $\omega_L$ are absorbed by the crystal and emitted at a number of shifted frequencies $\omega_i$ corresponding to creation of phonons. The likelihood for such an event to occur is described by second-order perturbation theory and can be expressed in terms of Raman susceptibility tensors. \cite{BorHua} The momentum carried by the photons is assumed to be small, and therefore only zone-centered phonons are considered.

In the context of density-functional perturbation theory, the Raman susceptibility tensors can be expressed as a sum over third derivatives of the total energy with respect to the electric field and phonon displacements, and in the case of longitudinal-optical phonons, a contribution arising from the generated macroscopic electric field. For details we refer to the papers by Gonze, Lee, and Veithen. \cite{Gon97,GonLee97,VeiGon05} The tensors were computed within the LDA using the \textsc{abinit} software. \cite{GonBeuCar02} We employed a plane-wave kinetic-energy cutoff of 30 hartree and sampled the Brillouin zone with a $10 \times 10 \times 10$  Monkhorst-Pack grid. Norm-conserving Troullier-Martin pseudopotentials were employed. \cite{TroMar91} For each volume the atomic positions were relaxed until the forces were less than 5\,meV/\AA. To enable comparison with the unpolarized experimental spectra, we averaged the calculated intensities over the individual polarizations and crystal directions.

\subsubsection{Wannier functions}
Wannier functions provide a powerful tool to visualize and analyze bonding in extended systems.\cite{SilMar98} A general transformation between the Bloch and Wannier representations is given by
\begin{align}
 w_n\left(\mathbf{R}\right) = \int_{\text{BZ}} \sum_m U^{(\mathbf{k})}_{mn}
     e^{-i\mathbf{k} \cdot \mathbf{R}} \psi_{m\mathbf{k}}\left(\mathbf{r}\right)
     d\mathbf{k}
   \label{eq:wann}
\end{align}
where $U^{(\mathbf{k})}_{mn}$ is a unitary matrix. By defining a measure of the spatial spread functional of the Wannier functions
\begin{align}
  \Omega=\sum_n 
    \left[
    \braket{w_n(\mathbf{0})}{r^2}{w_n(\mathbf{0})} - 
    \braket{w_n(\mathbf{0})}{\mathbf{r}}{w_n(\mathbf{0})}^2 
   \right],
\end{align}
which can be discretized in reciprocal space, Marzari and Vanderbilt \cite{MarVan97} were able to devise an algorithm to obtain so-called maximally localized Wannier functions (MLWFs). This algorithm only requires the overlap between the cell-periodic Bloch functions of neighboring $k$-points
\begin{align}
 M_{mn}^{\mathbf{k},\mathbf{b}} = \ibraket{u_{m\mathbf{k}}}{u_{n\mathbf{k+b}}}.
\end{align}

The MLFWs were constructed by interfacing \textsc{vasp} with the \textsc{Wannier90} program package. \cite{MosYat08} To compute $ M_{mn}^{\mathbf{k},\mathbf{b}}$ we used the earlier preconverged charge densities to generate Bloch functions on a $6 \times 6 \times 6$ Monkhorst-Pack grid. For the iterative minimization to reach the global minimum (with real-valued MLWFs) it is important to provide suitable initial guesses and their projections onto the Bloch states. These were calculated in real space using the soft part of the PAW wave functions. All MLWF images presented in the following were rendered by using the \textsc{POV-RAY}\cite{povray} software.

\section{Results}
\label{sec:results}

\subsection{Raman spectra}
\label{sec:raman}

 \begin{figure}
   \includegraphics[width=1.0\columnwidth]{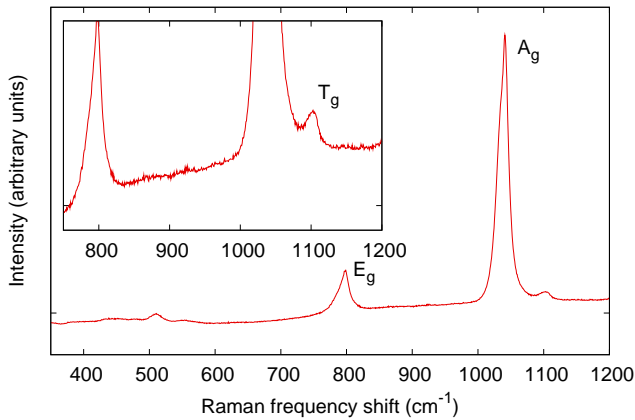}
   \caption{(Color online) Measured Raman spectra obtained after reaction between Pd and N$_2$. Spectrum obtained in experimental Run 1 at $\sim$ 57\,GPa. The inset shows in more detail the frequency range from 750\, to 1200\,cm$^{-1}$. Peaks are indexed assuming pyrite \PdN. The low-frequency peaks are due to nitrogen.
   \label{fig:spectraexp}}
 \end{figure}

Upon heating the sample, a reaction occurred above approximately 58\,GPa. This was indicated by the appearance of two new and relatively intense peaks in the Raman spectrum (\Ag\ and \Eg\ in \fig{fig:spectraexp}).\footnote{The widths and line shapes, and also the intensities, seemed often dependent on the position of the probed area and the experimental run (possibly related to the local stress state and amount of synthesized material, respectively). We conservatively assumed single Lorentzian line shapes in combination with linear or quadratic backgrounds when fitting the spectra.} A third weak feature on the high-frequency side of the most intense peak was observed (\Tg\ in \fig{fig:spectraexp}). We obtained a rough estimate of the synthesis temperature on the basis of Stokes to anti-Stokes ratios of the new peaks at the temperature at which they appeared. For the most intense peaks at 800\,cm$^{-1}$ and 1050\,cm$^{-1}$, we found values of 814\,K and 890\,K, respectively. According to group theoretical arguments, the Raman active modes of the pyrite structure are A$_{\text{g}}$, E$_{\text{g}}$, and T$_{\text{g}}$. \cite{bilbao} There are three distinct \Tg\ modes, and therefore five modes that can in principle be observed. Here, we surmise that two \Tg\ modes are missing in the experimental spectrum due to inherent weakness or proximity to one of the more intense modes. This is similar to the case of pyrite-\PtN\ where only four modes were observed and a weak \Tg\ mode was conjectured to be masked by one of the intense \Eg\ or \Ag\ modes. \cite{CroGon06a,YouMon06} 

\begin{figure}
  \includegraphics[width=0.5\columnwidth,trim = 0.1in 0.0in 0.0in 0.0in, clip]{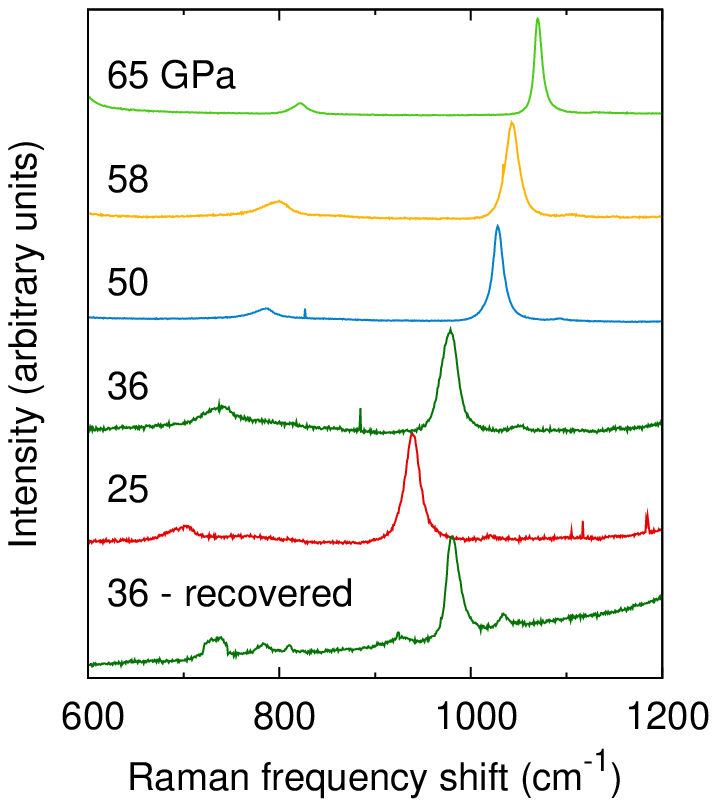} \nolinebreak
  \includegraphics[width=0.5\columnwidth,trim = 0.1in 0.0in 0.0in 0.0in, clip]{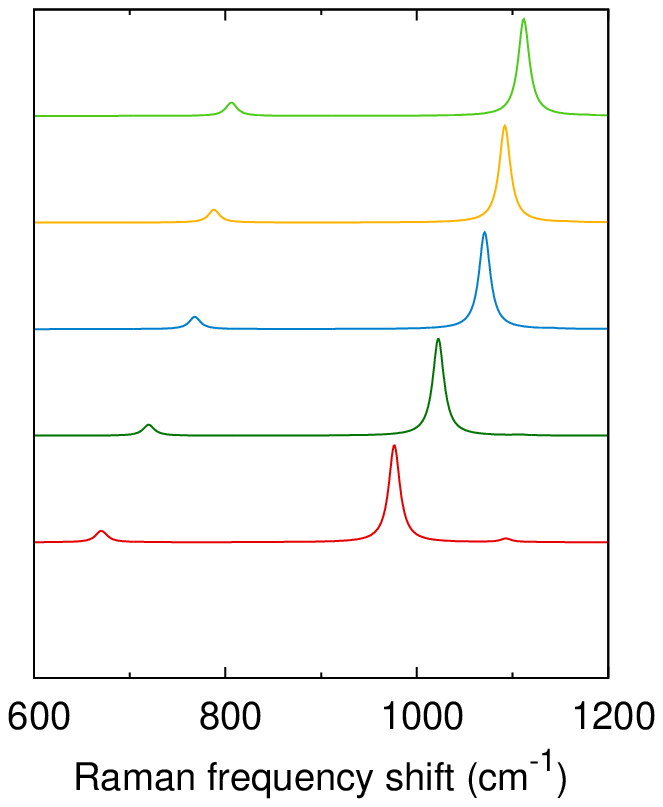} \nolinebreak
  \caption{(Color online) Raman spectra for a range of pressures. Left panel: experimental unprocessed data from Run 2. The lowest spectrum is obtained from the recovered sample after increasing pressure to 36\,GPa from a minimum of 11\,GPa. Right panel: calculated spectra (LDA). The spectra are normalized to maximum peak height.}
  \label{fig:spectra}
\end{figure}

Several Raman spectra were recorded at room temperature as a function of decreasing pressure. Figure \ref{fig:spectra} shows the experimental as well as the calculated Raman spectra of \PdN\ for a representative set of pressures. In general, the agreement is very good over the entire range, confirming the dominance of the \Ag\ and the \Eg\ modes in pyrite-\PdN.

Below 18\,GPa we did not obtain any convincing spectra. However, a type of spectrum characteristic of \PdN\ reappeared upon increasing the pressure from a minimum value (specifically from 11 to 36\,GPa, see \fig{fig:spectraexp}). \footnote{Actually the spectrum is not identical, see the bottom spectrum in \fig{fig:spectraexp}. The mode identified as \Tg\ is located at a lower frequency than might be expected, by approximately 15\,cm$^{-1}$, while the \Eg\ has a clearly different line shape perhaps also centered at a slightly lower frequency. There are also additional features around 800 and 925\,cm$^{-1}$ of unknown origin. In additional experiments visual observation at pressures below 10\,GPa at room temperature showed that the previously opaque sample appeared to become relatively transparent.} In an additional third experiment, very weak spectra were observed at room temperature down to 14\,GPa while  room temperature x-ray diffraction spectra were not observed below 13\,GPa. This strongly suggests that \PdN\ decomposes below this pressure.

\begin{figure}
  \setlength{\myhgt}{2.2in}
  \includegraphics[height=\myhgt]{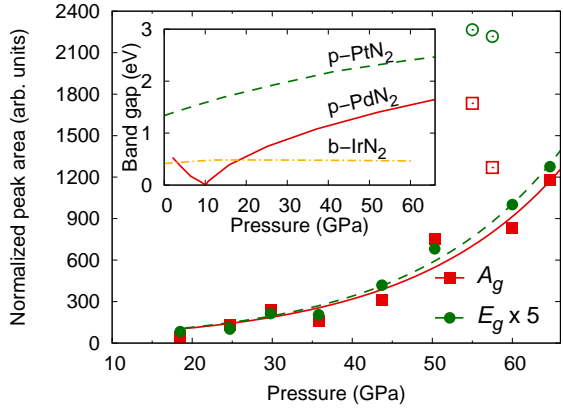}
  \caption{
    (Color online)  Experimental Raman intensities of the \Ag\ (squares) and \Eg\ (circles) modes as a function of pressure. The solid lines have been fitted to the filled data points only. 
  The inset shows the calculated band gaps of pyrite-\PdN, pyrite-\PtN, and baddelyite-\IrN. The band gaps of \PtN\ and \IrN\ are underestimated since they were calculated using a GGA exchange-correlation functional.
  }
  \label{fig:ramint}
\end{figure}

While the \PtN\ and \IrN\ compounds in our earlier experiments \cite{CroGon06a, CroGon08} exhibited Raman spectra whose intensities were not noted to be strongly pressure dependent, the relatively intense room-temperature Raman spectrum of \PdN\ weakens significantly as the pressure is lowered. We attempted to quantify this positive pressure dependence in \PdN\ by keeping the laser power reaching the diamond-anvil cell constant.\footnote{\label{ref:comment} We assume outlier points (open symbols in \fig{fig:ramint}) are due to not returning to exactly the same region of the sample after each pressure adjustment.} 
The resulting pressure dependence of the normalized spectral intensities, i.e., the Lorentzian areas for the \Ag\ and \Eg\ modes normalized by acquisition time is displayed in \fig{fig:ramint}. It can be well described by an exponential function. 

This behavior is indicative of a significant increase in the electromagnetic screening of the incident laser beam by the crystal as the pressure is reduced. In the inset of \fig{fig:ramint}, we show the HSE06 band gap of \PdN\ as a function of pressure. It clearly narrows as the pressure is reduced, leading to an exponential increase in the number of carriers excited across the gap, which in turn decreases the volume penetrated by the laser light. The theoretical 0\,K band gap vanishes at about 11\,GPa, leading to metal-like optical response of the material. The inset in \fig{fig:ramint} also shows the pressure dependence of the GGA band gaps of the two other insulating PM nitrides synthesized to date, \PtN\ and \IrN. In contrast to \PdN\, Raman spectra of these compounds have been measured over the entire experimental pressure range. This is consistent with the calculated band gaps which are always nonzero at non-negative pressures.


\begin{figure}
   \includegraphics[width=1.0\columnwidth]{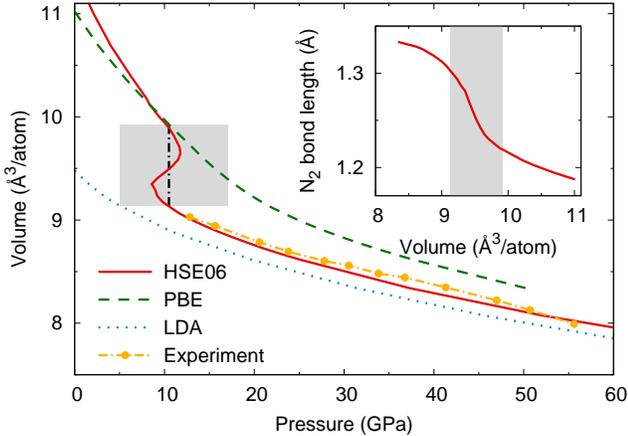} 
   \caption{(Color online) Equation of state as obtained from theory (HSE06, red solid, PBE, green dashed, and LDA blue dotted) and experiment (Ref. \onlinecite{CroGon08}) (yellow dashed-dotted) state. The inset shows the nitrogen bond length as a function of atomic volume and the gray areas indicates the two-phase region.}
   \label{fig:eos}  
\end{figure}

\begin{figure}
  \setlength{\myhgt}{2.2in}
  \includegraphics[width=\columnwidth]{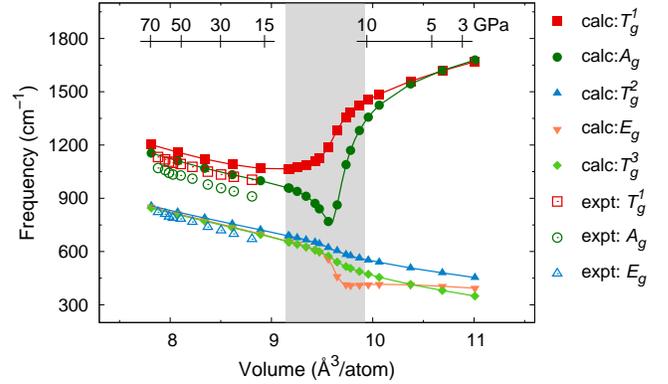}
  \caption{
    (Color online) Experimental and theoretical HSE06 Raman frequencies as a function of volume. The corresponding pressures as obtained from the equation of state are shown at the top of the figure. The gray area indicates the two-phase region (see \fig{fig:eos}).}
  \label{fig:ramfreq}
\end{figure}

\subsection{Equation of state and two-phase region}
\label{sec:eos}

Figure~\ref{fig:eos} compares equation-of-state (EOS) data from x-ray diffraction  measurements\cite{CroGon08} as well as LDA, GGA, and HSE06 calculations. The HSE06 calculations are in excellent agreement with the available experimental data which end at 13\,GPa. Below this pressure the HSE06 calculations show a two-phase region in which the EOS is multivalued corresponding to an isostructural transition within the pyrite structure.  As shown by the inset in \fig{fig:eos}, the transition is accompanied by a dramatic change in the nitrogen dimer bond length, which suggests a conversion from single to triple bonds as the pressure is decreased. We note that a change in bonding characteristics is also obtained by employing the LDA or GGA XC functionals (see discussions in Refs.~\onlinecite{YouMon06, CheTse10}). However, these two functionals do not yield a two-phase region. 

\begin{figure}
   \includegraphics[width=1.0\columnwidth]{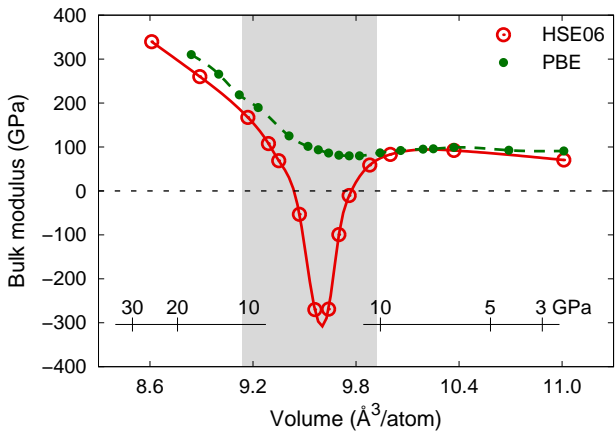} 
   \caption{(Color online) Bulk modulus as obtained from theory (HSE06, red solid, and PBE, green dashed). The gray area indicates the two-phase region.}
   \label{fig:elconst}  
\end{figure}

In the previous section we have related the disappearance of the Raman spectrum of palladium nitride at low pressure to the closing of the band gap. As can be seen in the inset of \fig{fig:ramint}, the band gap as a function of pressure is V shaped: it closes with increasing volume but reopens upon further expansion. If the electronic structure undergoes an abrupt transition in this region, one would expect a number of physical properties to exhibit unusual behavior. In fact, the HSE06-calculated Raman frequencies change very rapidly as seen in the gray area in \fig{fig:ramfreq}. In particular, the \Ag\ mode, which corresponds to the stretch mode of the nitrogen dimers in the pyrite lattice, first sharply drops and subsequently rises with pressure. Even more dramatic effects are observed in the long wavelength-acoustic phonon modes. In particular, the bulk modulus shown in \fig{fig:elconst} becomes negative in this region which provides a simple microscopic explanation for the decomposition below 13\,GPa observed in experiments. 

It needs to be emphasized that unlike the HSE06 results the LDA and GGA data predict a transition that involves no latent heat, a positive bulk modulus (\fig{fig:elconst}), and no unstable phonon modes.\cite{CheTse10} These XC-functionals cannot explain the experimental observations which implies that the isostructural phase transition in \PdN\ involves subtle electronic effects. 
In this context, it is also important to point out that for pyrite-\PtN\ we observed a shortening of the bond length at {\em negative pressures}. Furthermore, the relation between volume and pressure is single-valued at all pressures, regardless of the XC functional (LDA, GGA, and HSE06). This provides a consistent and sensible explanation for the experimental observation that \PdN\ decomposes under pressure whereas \PtN\ does not.


\subsection{Electronic structure of PdN$_2$}

\begin{figure}
\includegraphics[trim = 0.07in 0.0in 0.15in 0.0in, clip,height=2.2in] {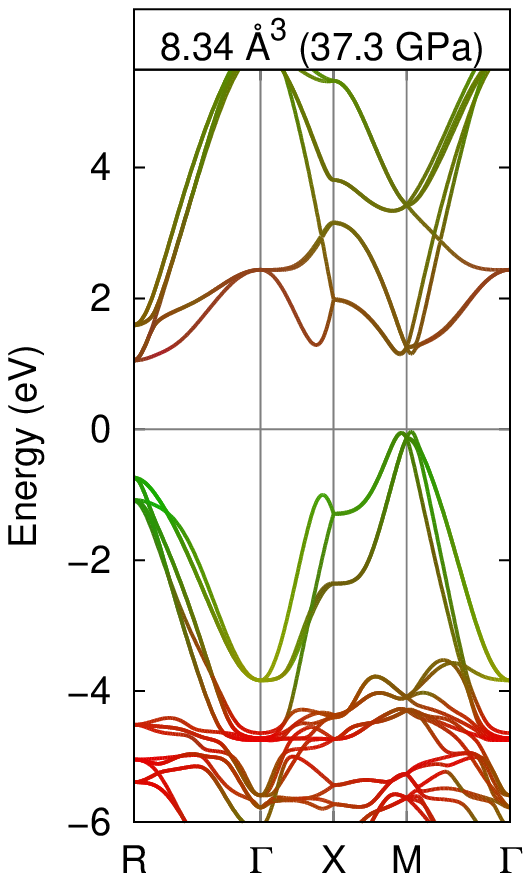} \nolinebreak
\includegraphics[trim = 0.45in 0.0in 0.15in 0.0in, clip,height=2.2in] {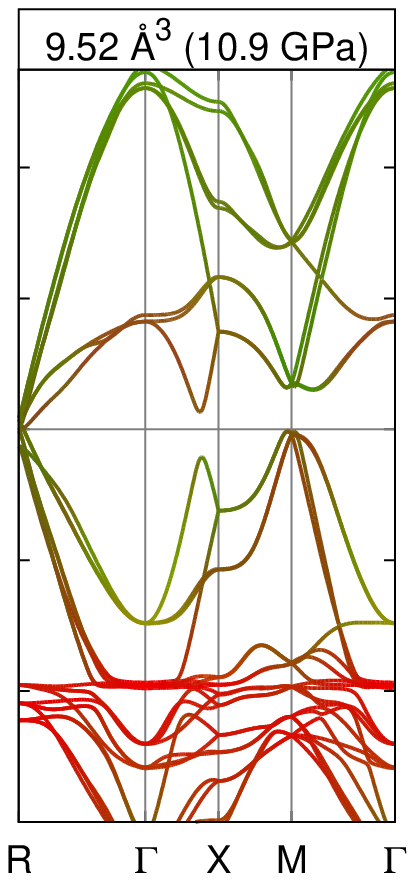} \nolinebreak
\includegraphics[trim = 0.45in 0.0in 0.15in 0.0in, clip,height=2.2in] {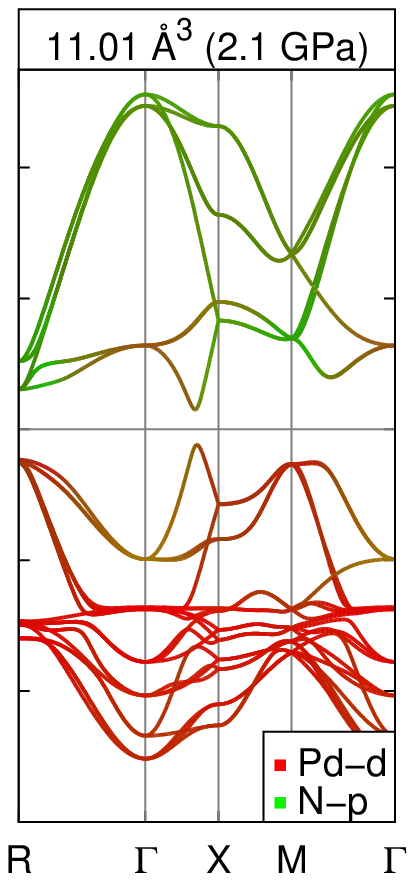} \nolinebreak
   \caption{(Color online) Orbital-projected band structures of \PdN\ at three representative pressures. The line color indicates the relative admixture of Pd-$d$ (red) and N-$p$ (green) states. Note the band crossing at the R point in the middle panel.
   }
   \label{fig:bnd}
\end{figure}

The band structure of pyrite-\PdN\ is strongly dependent on volume. This is illustrated in \fig{fig:bnd} where we display the orbital-projected HSE06 band structure at three representative pressures. At elevated pressures (left panel in \fig{fig:bnd}) the valence- and conduction-band edges are dominated by N $p$ and Pd $d$ states, respectively. At these pressures, \PdN\ is an insulator with a direct band gap at the M point. When the lattice is expanded, the band gap decreases until the two-phase region is reached. Here, the system becomes metallic with a small overlap of bands around the R point. At even larger volumes, after the transition has taken place, the system again becomes insulating with a direct band gap positioned at two thirds along $\mathrm{\Gamma - X}$ (right panel in \fig{fig:bnd}). The orbital character of the band edges is now reversed with respect to the high pressure case.

\begin{figure}
   \includegraphics[width=1.0\columnwidth,clip=true,trim= 0 7mm 0 0]{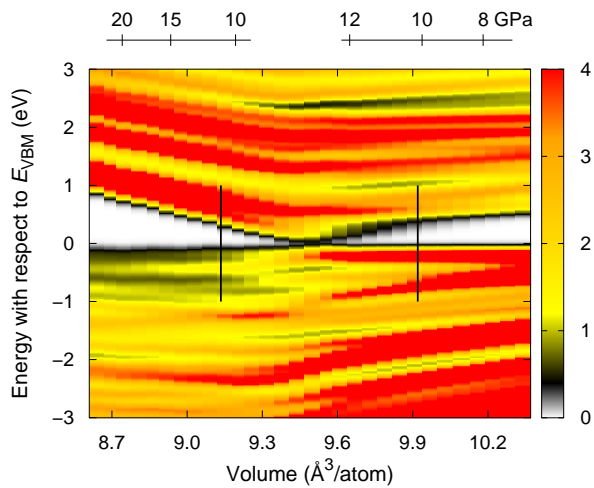}
   \caption{(Color online) Partial density of states (arbitrary units) for Pd-$d$ states as a function of volume and energy. The vertical bars indicate the two-phase region.
   }
   \label{fig:docc}  
\end{figure}

The transition in the electronic structure that accompanies the isostructural transformation is also illustrated in \fig{fig:docc},  which shows the density of $d$ electrons projected onto spheres of radius 1.4\,\AA\ around Pd atoms as a function of energy and volume. The white segments in \fig{fig:docc} approximately depict the V-shaped band gap as a function of volume with a minimum at 9.5 {\AA}$^3/$atom. 
\footnote{We note that the $k$-point sampling used to produce the projections is not sufficiently dense to capture the valence and conduction band extrema.Hence the band gap in this figure is overestimated.}
At this volume a massive transition in the character of the electronic states takes place as the Pd-e$_\text{g}$ states migrate across the band gap. In the following section, we will show that this migration is directly related to the conversion from single to triple nitrogen bonds alluded to above.

\begin{figure}
  \centering
  \includegraphics[width=\linewidth]{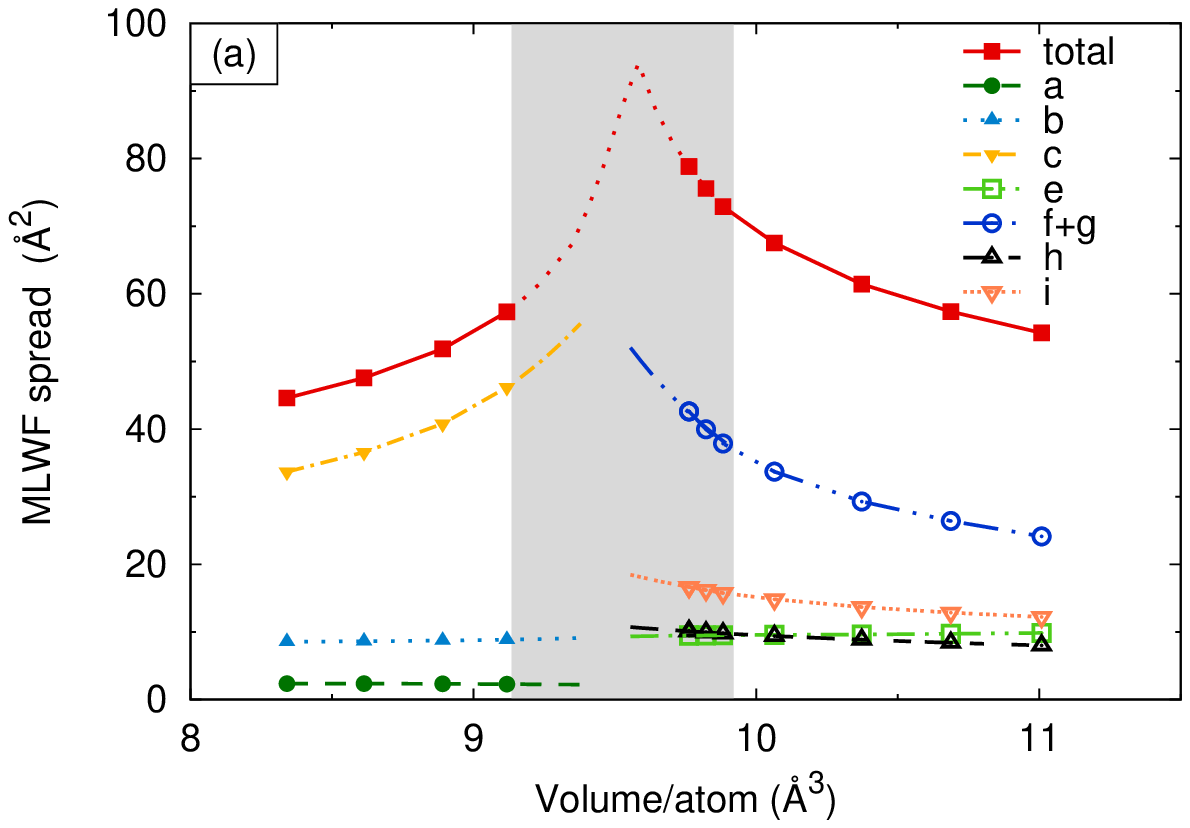} \\
  \includegraphics[width=\linewidth]{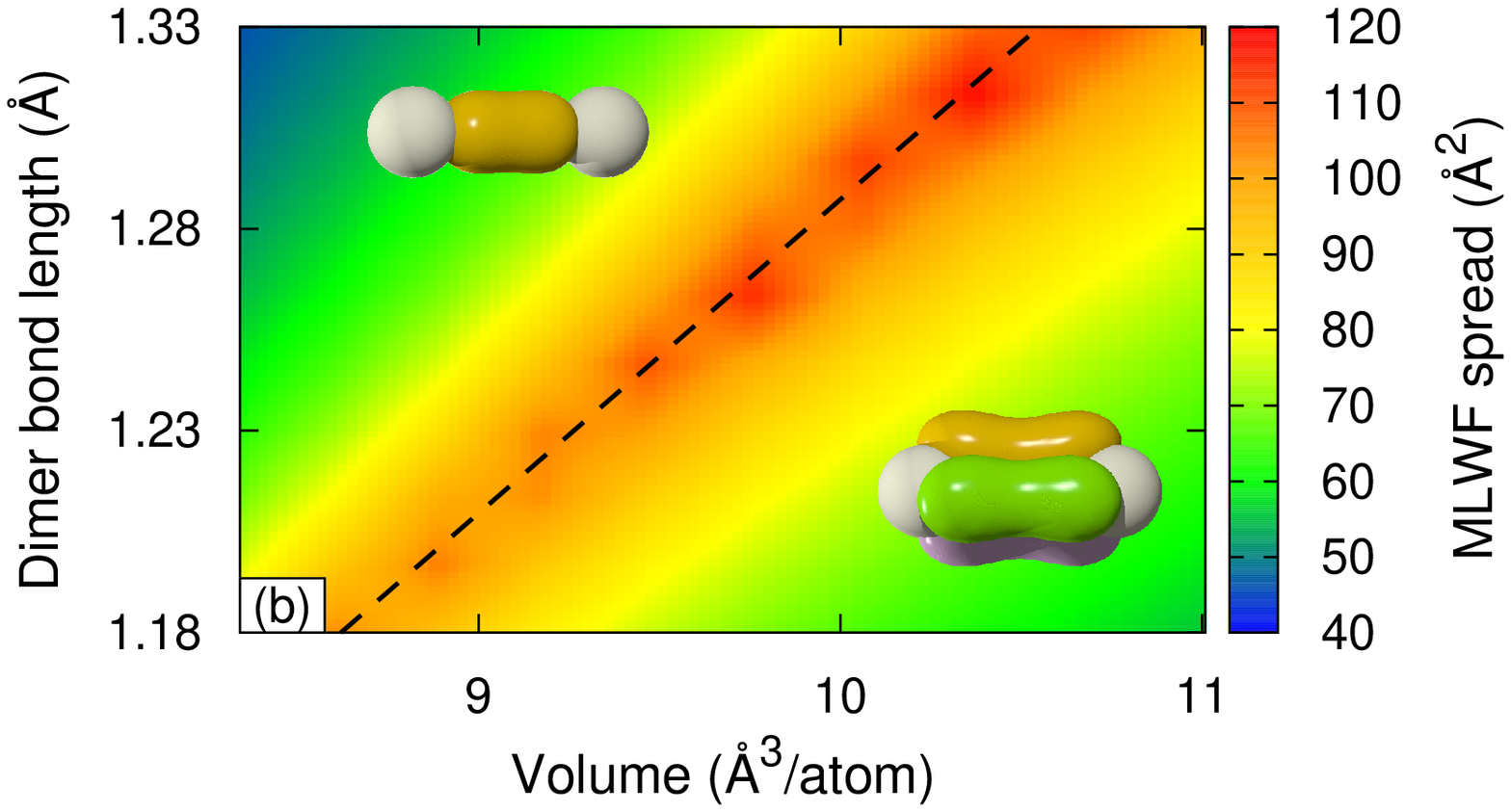} 
  \caption{
    (Color online) (a) Spread of occupied MLWFs. The red solid line indicates the sum of all occupied MLWFs. The letters in the legend correspond to the MLWFs depicted in \fig{fig:wann}. (b) Total spread of occupied the MLWFs as a function of volume and nitrogen dimer bond length. The dashed line separates the nitrogen single and triple bonded regions.}
  \label{fig:spread}
\end{figure}

\begin{figure*}[]
  \includegraphics[width=\linewidth]{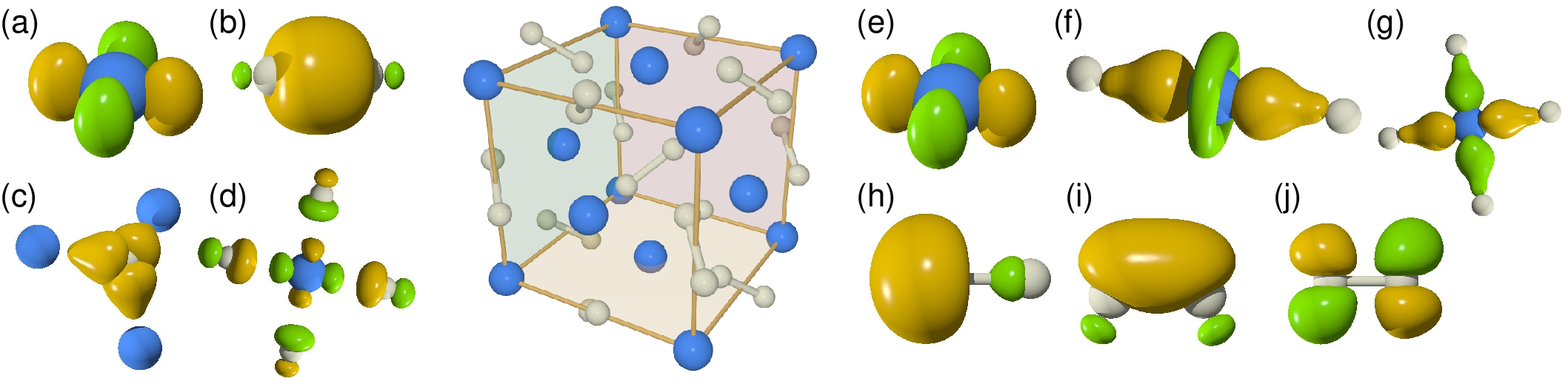}
  \caption{(Color online) (Center) Pyrite-\PdN\ conventional cell. [(a)-(j)] Isosurfaces of MLWFs orbitals, (a)-(c), and (e)-(i) correspond to {\em occupied} MLWFs in the high- and low-pressure regions, respectively. (d) and (j) correspond to the lowest {\em unoccupied} MLWFs. The green and orange surfaces denote positive and negative values, respectively. (a) $t_{2g}$-like states (three per Pd), (b) N-N single bond (one per dimer), (c) Pd-N bond (three MLWFs per nitrogen), (d) Pd $e_{g}$-like states (e)  Pd $t_{2g}$-like states (three per Pd), [(f) and (g)] Pd $e_g$-like states (one each per Pd), (h) N $s$ derived state (one per nitrogen), (i) N-N triple bond (three per dimer, each rotated 120$^\circ$ about the N-N axis), and (j) N$_2$-$2p\pi^*$-like states.} 
  \label{fig:wann}
\end{figure*}

\subsubsection{Wannier function analysis}
To further elucidate the nature of the electronic structure, we have performed an analysis of the MLWFs as a function of volume. In \fig{fig:spread}(a), we show the total spread of the occupied MLWFs (red solid line). The total spread rises from its initial value at elevated pressures until the system reaches the two-phase region. In the metallic phase, no separation of occupied and unoccupied states can be made. We can, however, construct approximate MLWFs by considering all Bloch states up to the Fermi-level as fully occupied. This is justified since in this region there is only a small volume in $k$-space around R where valence and conduction bands overlap. The spread obtained in this fashion rises sharply to a cusp [red dotted line in \fig{fig:spread}(a)]. At sufficiently large volume, the system becomes insulating again and the spread continues to decrease. Hence, the MLWF spread reproduces the behavior expected for a system undergoing an insulator-metal-insulator transition. 
As will be discussed in more detail below, the cusp in the MLFW spread functionals separates the regions of single and triple bonded nitrogen.

To fully characterize the electronic structure, we display the occupied and lowest unoccupied MLWFs in \fig{fig:wann} at high and low pressures. All occupied MLFWs were optimized simultaneously, while the conduction band MLWFs were analyzed separately.

\subsubsection{High-pressure region}
The occupied MLWFs in the high-pressure region can be classified as follows. While the site symmetry of a Pd ion in pyrite is strictly S$_6$, for the present purpose it can be approximated as cubic. In this case the five Pd \emph{d} orbitals (not counting spin degeneracy) split up into two $e_g$ and three $t_{2g}$ orbitals, the latter being lower in energy. This is verified by the MLWF-analysis as we obtain three occupied $t_{2g}$ states per Pd [\fig{fig:wann}(a)] and two unoccupied $e_g$-like states per Pd [\fig{fig:wann}(d)]. The latter MLWFs also have a significant contribution from N-$p_z$ orbitals. The nitrogen dimers are singly bonded via the N$_2$-2$s\sigma$ orbitals in \fig{fig:wann}(b). Palladium and nitrogen are bonded via the N-$sp$ hybridized orbitals shown in \fig{fig:wann}(c). These consist to a large extent of states derived from the N$_2$-$2p\pi^*$ orbitals.

\subsubsection{Low-pressure region}
At pressures below the transition, the Pd t$_\text{2g}$ orbitals are still occupied, and are joined by the Pd $e_g$ states [\fig{fig:wann}(f) and \ref{fig:wann}(g)] which here constitute the Pd-N bonding. The transition to triply bonded nitrogen dimers is now obvious: we obtain one occupied N $s$-like state [\fig{fig:wann}(h)] for each nitrogen, three identical MLWFs per dimer rotated 120$^\circ$ about the N-N axis [\fig{fig:wann}(i)]. Finally, the unoccupied MLWFs can be identified as N$_2$-$2p\pi^*$ anti-bonding states. This combined description of the N-N bond is analogous to the standard picture of an isolated triply bonded nitrogen dimer but modified to reflect the threefold rotation axis along the N-N bond.

\subsubsection{Single vs. triple bonded nitrogen}
In \sect{sec:eos} we have shown that during the course of the isostructural transition, the bond length of the nitrogen dimers decreases sharply. The Wannier function analysis discussed in the previous paragraphs has established that in the high- and low-pressure regions the nitrogen dimers are single [\fig{fig:wann}(b)] and triple bonded [\fig{fig:wann}(i)], respectively. We will now explore how the character of the N-N bond depends on volume and dimer separation. The total occupied MLWF spread shown in \fig{fig:spread}(b) displays a maximum in the metallic region along a straight line in the $V-d$ plane. Inspection of the occupied Wannier functions reveals that this line also separates the nitrogen single and a triple bonded regions. Based on the N-N separation, Chen {\em et al.} \cite{CheTse10} concluded that the nitrogen dimers in \PdN\ undergo a transition from single via double to triple bonded with decreasing pressure. Figure~\ref{fig:spread}b, however, clearly shows the character of the N-N bond to depend on both volume {\em and} dimer separation. Also, we do not observe double bonded nitrogen dimers.

\subsubsection{Band anticrossing}
\begin{figure*}
\centering
  \includegraphics[height=2.6in,trim = 1.0in 0.0in 0.0in 0.0in, clip]{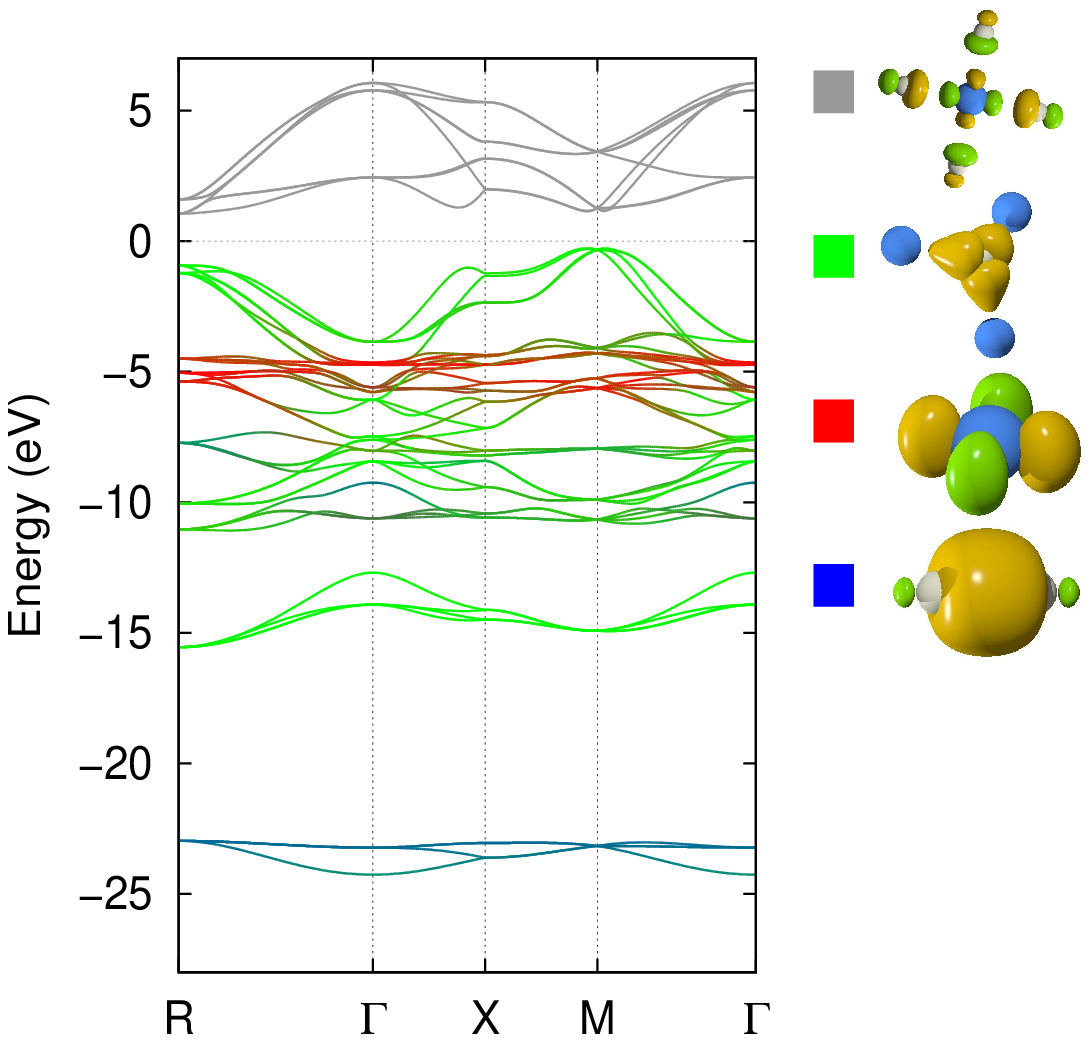}
  \includegraphics[height=2.6in,trim = 1.0in 0.0in 0.0in 0.0in, clip]{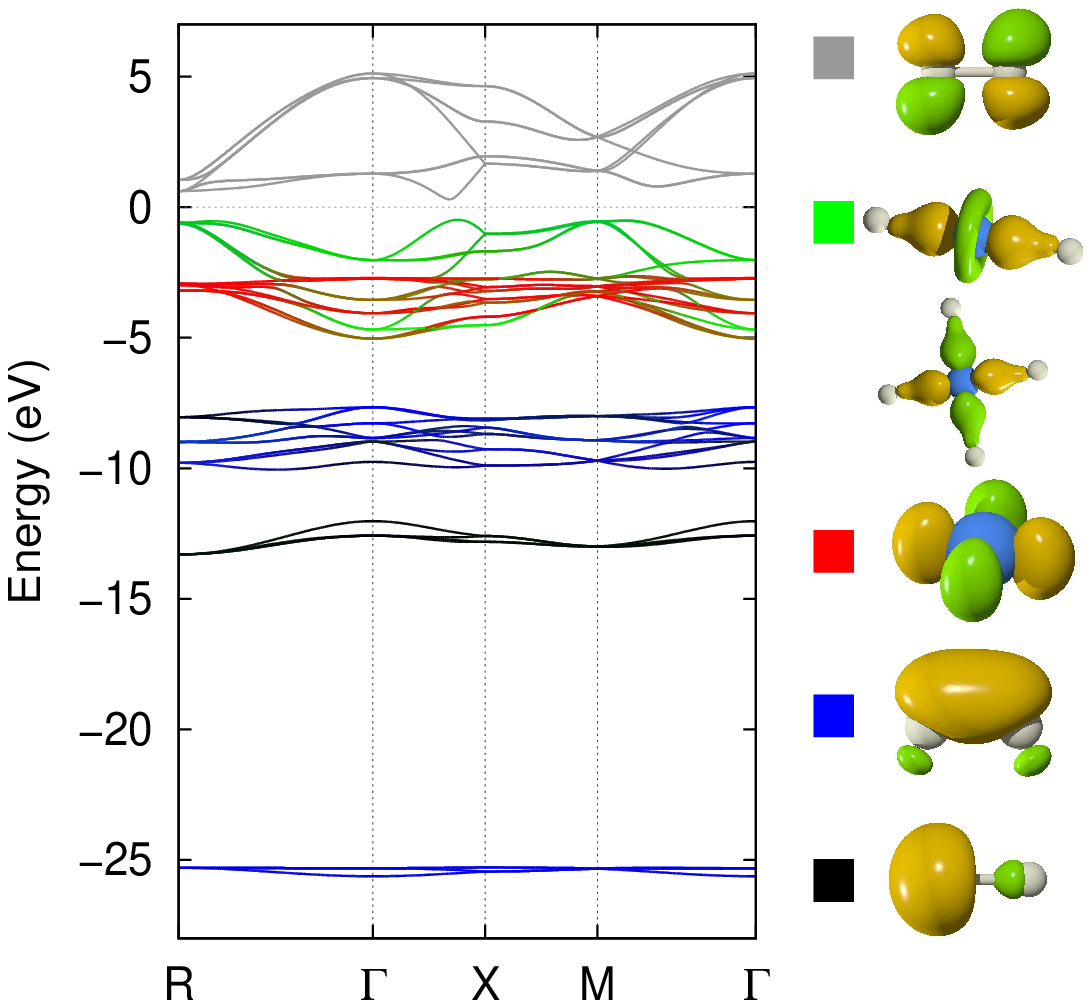}
   \caption{(Color online) Band structures for the occupied states in the high- (left panel) and low-pressure cases (right panel), respectively. The line color indicates the projection of the eigenstates onto the MLWFs.}
  \label{fig:wanproj}
\end{figure*}

Knowledge of the rotation matrices  $U^{(\mathbf{k})}_{mn}$ in \eq{eq:wann} enables us to decompose the band structure to show the relative contribution of each Wannier center. The results for the high- and low-pressure cases are displayed in \fig{fig:wanproj} which complements the site and angular momentum projected band structures (\fig{fig:bnd}). In both cases, the MLWFs corresponding to the N-N bonds are associated with bands far below the valence-band edge. At high pressure the valence- and conduction-band edges are dominated by Pd $e_g$ states and N$_2$ $2p\pi^*$ states, respectively. The band character is reversed in the low-pressure phase without any significant rearrangement of the bands. However, as the states associated with the band edges belong to the same irreducible representation, band anticrossings are observed both as a function of reciprocal lattice vector and volume. 

The eigenvector exchange associated with the band anticrossing first explains the closing and re-opening of the band gap. Second, it leads directly to the migration of Pd-e$_\text{g}$ and N$_2$-$2p\pi^*$ states across the gap as a function of volume. Third, the appearance of N$_2$-$2p\pi^*$ states in the valence band in the high-pressure phase explains the increase in the nitrogen dimer bond length and the transition from triple to single bonds.

\section{Discussion}
\label{sec:discussion}

In \sect{sec:raman}, using Raman spectroscopy it was shown that a compound of Pd and N is synthesized in a diamond-anvil cell at pressures above 58\,GPa. 
Convincing Raman spectra were not observed below 18\,GPa at room temperature (\fig{fig:spectra}). In an earlier study,\cite{CroGon08} we observed that the x-ray diffraction signal due to the face-centered-cubic (fcc) Pd metal sublattice disappeared at pressures below 13\,GPa. In conjunction, these observations suggest that \PdN\ synthesized at high pressures decomposes if the pressure is released.
The very good agreement between experimental data and Raman spectra calculated for pyrite-\PdN\ furthermore confirms the previous structural assignment.\cite{CroGon08, Aberg08}

Unlike any of the other PM nitrides investigated so far, the Raman intensities for \PdN\ decrease exponentially with pressure (\fig{fig:ramint}). This intriguing behavior can be understood by inspection of the calculated band gap, which displays a positive approximately linear pressure dependence in the respective range closing at 11\,GPa. The band-gap narrowing with decreasing pressure leads to an exponential increase in the free carrier concentration and dielectric screening, causing an exponential decay of the Raman intensities. The strong and non-monotonic volume dependence of the band gap foreshadows the unusual properties of this compound near the decomposition point. 

As shown in Ref.~\onlinecite{Aberg08}, \PdN\ and \PtN\ are qualitatively similar with respect to the pressure dependence of formation enthalpies. Nonetheless, \PtN\ has been successfully recovered to ambient conditions while \PdN\ decomposes already at 11--13\,GPa. In \sect{sec:eos}, we investigated the EOS to resolve the origin of this difference. Above the decomposition pressure, excellent agreement was obtained between experimental data and calculations using the HSE06 hybrid XC functional, which mixes the conventional PBE XC functional with 25\%\ screened exchange. Since the symmetry constraints imposed in the calculations prevent the decomposition of the compound, the EOS can be mapped out below the decomposition pressure. The calculations reveal a first-order isostructural phase transition right in the vicinity of the experimental decomposition pressure. In the two-phase region (gray bar in \fig{fig:eos}), several properties become extremal: the band gap closes and reopens (\fig{fig:ramint}) and the Raman frequencies exhibit a pronounced dip (\fig{fig:ramfreq}). Most importantly, however, the bulk modulus becomes negative (\fig{fig:elconst}) which readily explains the experimentally observed decomposition as being triggered by a long-wavelength acoustic phonon instability. It is important to note that conventional XC functionals such as LDA or GGA do neither agree well with the experimental EOS nor do they yield a first-order phase transition associated with an instability. This clearly highlights the importance of electronic effects in understanding this transition.

As the first-order transition is isostructural, the two phases could be electronically distinct. In fact, a systematic Wannier analysis shows that on the high- and low-pressure side of the transition the nitrogen dimers are single and triple bonded, respectively [\fig{fig:wann}(b)]. Along the transition pathway, the total spread of occupied Wannier functions increases as the two-phase region is approached, diverges in the metallic transition region, and subsequently drops again. Also during the transition the characters of the valence- and conduction-band states are inverted corresponding to a band anti-crossing with eigenvector exchange (Figs.~\ref{fig:bnd} and \ref{fig:docc}). We find that both volume and the nitrogen dimer bond length determine the character of the N-N bond, rendering assignments based solely on the bond length unreliable (\fig{fig:spread}). Furthermore, based on the Wannier analysis the nitrogen dimers are never double bonded as speculated earlier.\cite{CheTse10}

\begin{figure}
  \centering
  \includegraphics[width=0.8\columnwidth]{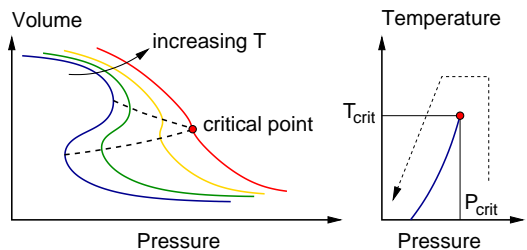} 
  \caption{(Color online)
    Schematic illustration of the relation between pressure and volume that gives rise to a first-order transition at low temperatures and becomes second order as temperature is increased. The inset exemplifies a possible corresponding phase diagram.
  }
  \label{fig:schematic}
\end{figure}

Finally, a discussion of the character of the isostructural transition is in order. A well-known example of an isostructural phase transition is the fcc-fcc transition in cerium.\cite{Bri48} At low temperatures, this transition is first order but as schematically shown in \fig{fig:schematic} the hysteresis diminishes with increasing temperature which eventually leads to the phase separation line terminating in a critical point. The occurrence of a critical point allows one to avoid the first-order transition by first increasing temperature, then pressure, and subsequently reducing temperature again as indicated by the dashed lined in \fig{fig:schematic}. The EOS shown in \fig{fig:eos} suggests that the isostructural transition in \PdN\ could also terminate in a critical point. While it is beyond the scope of the present work to answer this question, such behavior would provide an intriguing pathway to recover the \PdN\ to ambient conditions.\\[6pt]

\begin{acknowledgments}
This work was performed under the auspices of the U.S. Department of Energy by Lawrence Livermore National Laboratory in part under Contract No. DE-AC52-07NA27344, as well as being based on work supported as part of the EFree, an Energy Frontier Research Center funded by the U.S. Department of Energy, Office of Science, Office of Basic Energy Sciences under Award No. DE-SC0001057.
\end{acknowledgments}

\end{document}